\documentclass[conference]{IEEEtran}
\IEEEoverridecommandlockouts
\usepackage{cite}
\usepackage{amsmath,amssymb,amsfonts}
\usepackage{algorithmic}
\usepackage{graphicx}
\usepackage{textcomp}
\usepackage{xcolor}
\usepackage{comment}
\usepackage{mathtools}
\usepackage{xspace}
\usepackage{comment} % biblioteca de comentários
\usepackage{tikz}

\newcommand\copyrighttext{%
  \footnotesize This work has been accept  to the IEEE ISCC2020. Copyright 978-1-7281-8086-1/20/\$31.00~\copyright2020 IEEE.}
\newcommand\copyrightnotice{%
\begin{tikzpicture}[remember picture,overlay]
\node[anchor=south,yshift=8pt] at (current page.south) {\fbox{\parbox{\dimexpr\textwidth-\fboxsep-\fboxrule\relax}{\copyrighttext}}};
\end{tikzpicture}%
}

\newcommand{\as}[1]{\textcolor{blue}{{\bf #1}}} %Aldri

\newcommand{\cm}[1]{\textcolor{red}{{\bf #1}}} %carlos
\newcommand{\al}[1]{\textcolor{brown}{{\bf #1}}} %Yan
\newcommand{\sep}{\hspace{8 mm}}
\def\BibTeX{{\rm B\kern-.05em{\sc i\kern-.025em b}\kern-.08em
\kern-.1667em\lower.7ex\hbox{E}\kern-.125emX}}

\definecolor{color1}{RGB}{198, 90, 103}
\definecolor{color2}{RGB}{175, 42, 48}
\definecolor{color3}{RGB}{0,128,128}
\definecolor{color4}{RGB}{179, 43, 59}
\definecolor{color5}{RGB}{128,0,128} %Yan
\definecolor{color6}{RGB}{4,159,255} %YanNovo

\usepackage[english,ruled,noline,linesnumbered,noend]{algorithm2e}

\title{%Consensus-based Cooperative Task Assignments
Managing Consensus-Based Cooperative Task Allocation for IIoT Networks
\\
}

\author{
\IEEEauthorblockN{
{\bf Carlos Pedroso\IEEEauthorrefmark{1}, }%\IEEEauthorrefmark{2}},
{\bf Yan Uehara de Moraes\IEEEauthorrefmark{1}, }%\IEEEauthorrefmark{2}},
{\bf Michele Nogueira\IEEEauthorrefmark{1}, }%\IEEEauthorrefmark{2}},
{\bf Aldri Santos\IEEEauthorrefmark{1}}}

\IEEEauthorblockA{\IEEEauthorrefmark{1}Wireless and Advanced Networks Laboratory (NR2) - UFPR, Brazil  \\
Emails: \{capjunior, yumoraes, michele, aldri\}@inf.ufpr.br %\\ 
}}

\IEEEaftertitletext{\vspace{-1.0\baselineskip}}

\begin{document}
\maketitle
%\IEEEpubidadjcol
\copyrightnotice
\thispagestyle{plain} \pagestyle{plain} % Comando para enumerar paginas

\begin{abstract}
Current IoT services include industry-oriented services, which often require objects to run % perform 
more than one task. However, the exponential growth of objects in IoT poses the challenge of distributing and managing task allocation among objects. One of the main goals of task allocation is to improve the quality of information and maximize the tasks to be performed. Although there are approaches that optimize and manage the dynamics of nodes, not all consider the quality of information and the distributed allocation over the cluster service. This paper proposes the mechanism CONTASKI for task allocation in IIoT networks to distribute tasks among objects. It relies on collaborative consensus to allocate tasks and similarity capabilities to know which objects can play in accomplishing those tasks. CONTASKI was evaluated on NS-3 and achieved 100\% of allocated tasks in cases with 75 and 100 nodes, and, on average, more than 80\% clusters performed tasks in a low response time.
%}
\end{abstract}


%\al{Lembretes: Harmonizarmos o texto para a voz ativa; dentro do possível usar 1a pessoa do plural; criarmos uma tabela de vicios de escrita}

\section{Introduction}
The Internet of Things (IoT) is a heterogeneous network where objects hold  various characteristics, such as identity, physical attributes, computational and sensing capabilities ~\cite{gubbi2013internet,botta2016integration}. For the vast majority of IoT objects, it is important to reduce the power consumption by communicating or making  certain tasks. Further, objects must evenly share their resources and cooperate to support a better network performance~\cite{khalil2019new}. Thus, objects that run a set of 
%\as{the same/a set of} 
functions can collaborate to allocate and realize different tasks~\cite{ghanbari2019resource}. 
\begin{comment}
Particularly, IoT is directly tied to the evolution of 
%a couple of / various %several/
different 
complex application domains, such as industrial processes, logistics, smart health, and smart cities. In this context, 
\end{comment}
In this context, Industrial Internet of Things (IIoT) has %recently 
received greater attention~\cite{xu2018industry} since it focuses on connecting different objects with numerous capabilities within an industry, enabling thus everyone to work in a synchronized and organized manner to perform tasks.

%\al{Este trecho abaixo está confuso.} \\

The task allocation among IoT objects is 
%becomes 
an alternative 
to deal with  %/to handle 
%to addressing 
the resource heterogeneity of objects that can perform more than one sensing 
activity in the network, like temperature measurement, environment monitoring, positioning among others. Industrial services as well as logistics, inventory control, and machinery synchronization can benefit from task allocation~\cite{zhong2017intelligent}. Thus, 
%This is because 
tasks can be allocated by evaluating capabilities that each object is capable of performing within the network~\cite{korsah2013comprehensive}. However, 
IoT is dynamic and typically has mobile and fixed nodes, and infrastructure varies with the interactions between objects. A range of objects have limited resources, low power, low processing and storage capacity, and loss of connection links~\cite{qiu2018can}. So dealing with this variation poses challenges regarding the quality of the data generated by the objects. The poor quality of information available makes it difficult to interpret by applications. Using object groupings contributes to network organization, decreased resource consumption, and the quality with which information is disseminated\cite{gielow2015data}. Thus, the task allocation problem is not trivial~\cite{colistra2014task} due to the size and configurations that an IIoT network can achieve~\cite{santos2019clustering}.

%Due to the size and configurations that an IoT network can achieve, the task allocation problem is not trivial~\cite{colistra2014task}.

%In particular,  %Particularly, 
Addressing 
%Handling/Dealing with 
resource allocation issues to get 
%achieve 
good performance for running 
%performing 
tasks among different 
IoT objects has been 
challenging%a major challenge
~\cite{aazam2018iot}. Factors, like 
%such as 
resource scarcity, object heterogeneity, and environmental dynamics usually struggle the task distribution management.  
%have made difficult managing properly % correctly
%task distribution. 
Resource allocation required at the devices becomes essential to the fairness of task distribution among IoT network participants, without 
the management of resource allocation multiple nodes will waste likely resources unnecessarily or even fail at crucial moments for role definition. A faulty distribution of tasks leads to flaws in the quality of the information generated by IoT objects, affecting services as data dissemination and consequently the applications. Besides, the efficiency of their maintenance must be ensured so that nodes can provide the quality of allocated services~\cite{fang2018earning}. Hence,
%Therefore, 
conscious resource management is needed to better utilize IoT objects, making everyone work according to their characteristics and functions.

\begin{comment}
Resource allocation \as{required/existing at the devices} becomes important in the fairness of task distribution among IoT network participants, without \as{it/ quem é ele?} multiple nodes 
will 
%may consume
\as{consume/waste} likely resources unnecessarily or even fail at crucial moments for role definition~\cite{ghanbari2019resource}. An \as{weak/wrong/mistaken}
%poor 
distribution of tasks leads to
%problems failure/gaps  % failure, fault, flaw, defect, shortcoming, rift, inadequacy,
flaws in the quality of the information %that is 
\as{generated by/come from} IoT objects, affecting 
%causing services such as data dissemination to be affected.
services as data dissemination. 
%Besides, an effective maintenance enables nodes can provide the quality of allocated services~\cite{fang2018earning}. 
\as{Besides, the efficiency of its maintenance must be ensured so that nodes can 
provide 
%guarantee 
the quality of allocated services~\cite{fang2018earning}.} 
\end{comment}

%Generally~\pl{news?application?} 
%Although 
%In wireless sensor networks (WSN)
Task allocation services have been 
%commonly 
studied 
extensively in wireless sensor networks (WSN), which usually treats it as resource allocation. In particular, task allocation collaborate 
%is employed 
to improve the life of the network~\cite{pilloni2017consensus}. With the rapid advancement of IoT networks, task allocation solutions have also become part of this research scope~\cite{ghanbari2019resource}. Among 
%the 
works that %seek to 
deal with the task allocation problem, many consider it through the use of object virtualization in task groups~\cite{khalil2019new} and distributed consensus~\cite{pilloni2017consensus,colistra2014task}. Object virtualization is used to assign tasks
according to the sensing competencies that each object has and its performance capacity. The main objective
%\as{objective/goal} 
is the optimization of tasks to save resources. As fair distribution of resources, distributed consensus applies based on the characteristics of each group of objects in the network. This approach is mostly applied to networks with a large number of participants. 
%Distributed consensus \as{is employed/ Voz passiva} in the equitable distribution of resources based on the characteristics
%of each group of objects present in the network, this approach is mostly \as{used/applied to} in networks with a large number of participants. 
However, these solutions do not consider the similarity relationship among objects and tasks to be performed, as well as the 
%definition of 
allocation based on the characteristics of the environment in which the objects are inserted. Though,
%Thus, 
it is 
%becomes 
crucial for IoT 
%that mechanisms be able 
to own mechanisms able to manage %\as{distribute/manage} 
task allocation among objects 
through their relationships and capabilities for more robust and conscious management of available network resources.

This paper 
%presents 
introduces 
a mechanism 
for supporting 
%to support 
the task allocation service among
%\as{between/among} 
objects into an IoT network, called CONTASKI (\emph{\textbf{CON}sensus Collaborative Based \textbf{TASK} Allocation for~\textbf{I}IoT}). It
%organizes 
arranges 
the network into similarity-based groups 
%to handle 
to 
handle 
%address  
the division of tasks to be allocated. 
%the allocation of tasks among the devices.
%the role of tasks allocated to devices 
CONTASKI 
makes use of 
%employs 
a distributed consensus strategy for decision making about 
%what is the better task distribution for a given service. 
%which distribution is best to use for a given service. 
the better tasks distribution for making a given service. 
%\as{
Evaluation on NS-3 simulator has shown that CONTASKI achieves  100\% of allocated tasks in certain cases with 75 and 100 nodes, and also, on average, more than 80\% clusters performed tasks, keeping thus the quality of the information disseminated by IIoT object.

This paper is organized as follows: Section~\ref{sec:rel} discusses the related work.  Section~\ref{sec:sys} defines the model and assumptions taken by CONSTASKI.
Section~\ref{sec:CONF}  describes the CONSTASKI components and their operation. Section~\ref{sec:ana} shows the evaluation methodology 
%to analyze the performance of the mechanism, 
and %accompanied by 
the results obtained. Section~\ref{sec:con} presents conclusions and future work.

\section{Related Work}
\label{sec:rel}

The provision of dynamic and distributed services 
aware of the resources and capabilities of IoT objects has been the focus of several works~\cite{khalil2019new,pilloni2017consensus,colistra2014task,pilloni2011deployment}. 
The adoption of means of task allocation management enables the maximization of resource usage  among network objects. However, 
coordinating the distribution of these resources entails challenges in their conduct, 
like 
%such as
%~\textit{(i)} 
assessing the capabilities of objects concerning  tasks to be performed,
%~\textit{(ii)} 
organizing the network and
%~\textit{(iii)} 
ensuring 
%the 
quality of the information~provided.

In~\cite{pilloni2017consensus}, virtual objects (VO) in an IoT Smart health network implement a decentralized strategy for allocating tasks 
where they negotiate with each other to reach a resource allocation consensus.
%that negotiated among them  to reach a resource allocation consensus.
%The authors in~\cite{pilloni2017consensus} proposed a new distributed algorithm for virtual objects (OV) in an IoT Smart health network. OVs can implement 
%\as{eles implementam ou pode implementar?} 
%a decentralized strategy for allocating tasks that negotiate with each other to reach a resource allocation consensus. T
%They extended the model of information that objects receive to include new features in a distributed scenario, including the quality of information that measures the characterization of the information. Although the work has followed a promising line, 
However, 
it does not evaluate the issues of objects having various capacities and types of interactions and also the influence of network size. In~\cite{khalil2019new}, it is proposed 
an~evolutionary algorithm based on a heterogeneity recognition heuristic to ensure greater stability and operational periods of tasks in an IoT network. The model creates 
%a 
collaboration between the functions of IoT objects based on the tasks to be executed and in the selected groups. The algorithm selects objects with energy levels above the average level of the other task group members. Thus, group's virtual objects can perform tasks and also reduce the energy consumption. 
But, 
%Though, %However, 
the model assumes that few objects are capable of performing all tasks defined by the network, and only two types of tasks are performed, restraining the use of the model to networks with multiple nodes and capabilities.

%In~\cite{colistra2014task}, the authors developed a consensus-based heuristic approach to task allocation. The model is a distributed method of task allocation in IoT with the primary purpose of fault tolerance.
In~\cite{colistra2014task}, 
%the authors developed 
a consensus-based heuristic approach to task allocation in IoT with the primary 
%purpose 
goal 
of fault tolerance 
%The algorithm uses the concept of task groups and objects. In each task group, objects can be selected as virtual and vice-virtual. 
makes use of 
%They use 
the concept of task groups and objects 
so that 
%in way that 
in each task group, objects can be selected as virtual and vice-virtual.
This model assists the best division of tasks between objects, in way that when a virtual object runs out of energy, the vice-virtual object takes on the duty of being a virtual object and the next in the list becomes the vice-virtual object of the corresponding task group. However, the model requires periodic~\textit{hello} message exchanges, 
being costly in 
%which 
%%represent 
%mean
%management messages. 
%This excessive message exchange represents 
communication, computing, storage, and power overload. In addition, 
the capabilities of nodes are not considered,
%they do not consider the capabilities of network nodes, 
which directly influence the distribution of tasks. In~\cite{kim2015efficient}, to deal 
%the authors proposed dealing 
with the task allocation issue in IoT, they consider that all objects cannot interact directly with one another, and make use of ~\textit{gateways} services to be responsible for managing this interaction. Thus, they have transformed the task allocation problem into an integration problem with a minimal degree variant in order to narrow the problem and thus can apply a genetic algorithm to reduce the time required to allocate tasks. However, 
%using 
the interaction manager ultimately limits relationships between nodes, and the information centralization model can 
cause a communication bottleneck depending on the network size.

In~\cite{pilloni2011deployment}, an algorithm 
decomposes sensor tasks into distributed tasks by taking the energy consumption for making each task and the feasibility of the solution in the assignment of tasks between the sensors. 
%the authors proposed an algorithm that decomposes sensor tasks into distributed tasks. The algorithm evaluates the energy consumption for the execution of each task and, thus, evaluates the feasibility of the solution in the assignment of tasks between the sensors. 
However, the model only considers %apply/consider/employ  
a centralizing entity in the distribution of functions, not considering a leader-based network organization for task distribution between nodes. Although effective in a dynamic network such as IoT may not be viable given the standalone scenarios. %Following the same line
In~\cite{jin2011intelligent}, it is proposed an algorithm that uses adaptive task mapping in sensors. It works in parallel with scheduling based on a genetic algorithm in which both work in real-time. The algorithms aim to extend network life by balancing the workload between sensors.
%presented an algorithm that uses adaptive task mapping in sensors. He works in parallel with scheduling based on a genetic algorithm in which both work in real-time. The algorithms aim to extend network life by balancing the workload between sensors. 
However, by centralizing the entire distribution of tasks, they overload the transmission channel, as well as inserting a delay in the delivery of messages, which end up compromising the synchronization of tasks execution.

\begin{comment}
Thus, we argue that the task allocation management 
should be  capable of acting in a collaborative and distributed way among objects, by linking tasks according to the capabilities of each object of 
the IoT network. These solutions, while~\cm{preserving}
%\as{preserving/saving} %the 
resources of objects, should not 
yield additional overhead, and 
usually meet characteristics of the IoT network and~\cm{the type of environment where it is located.}
%\as{environment?} 
%in which it operates.
\end{comment}

\section{System Model}
\label{sec:sys}

This section presents the 
%describes our 
%assumptions on the 
network, communication and task models. 
%As shown in Figure~\ref{fig:network-model}, 
The IIoT network %model %consists of 
%assumes 
takes into account
%an infrastructure divided into clusters, 
an infrastructure arranged into clusters, 
%which are 
composed of common nodes, leader nodes, and APs.
The task model
%\as{model/procedure/process} 
comprises the task description and its life cycle.
The communication model is responsible for keeping all the nodes and leaders connected. 

\begin{comment}

\begin{figure}[ht]
	\centering
	\includegraphics[width=1\linewidth]{images/modelo_de_rede.pdf}
	\caption{Network model}
	%\vspace{0.5cm}
	\label{fig:network-model}
\end{figure}

\end{comment}

\subsubsection{\textbf{Network model}}
%the IoT network comprehends a set of N nodes in an area $(X_{x}, Y_{y})$.
%Figure~\ref{fig:network-model} shows the network model which is
%consists of %handles 
An IIoT network 
%$N$ 
composed by a set of objects denoted by $N = \{ ob_{1}, ob_{2}, ..., ob_{n} \}$ in an area $(X_{x}, Y_{y})$. All objects have an unique identifier $ID$ to identify them in the network. The objects differentiates among themselves by their capabilities set $C = \{ c_1, c_2, c_3, ..., c_n \}$, processing power, energy and memory.
The objects are static and evenly distributed in an area of the network $N$ with coordinates $ \{ (x_1, y_1), (x_2, y_2), ..., (x_n, y_n) \} $.

\subsubsection{\textbf{Communication model}}

% MessageTypes{TaskDispatch, TaskAccept, LeaderRegister, CapabilityDissemination, LeaderToCluster}
Communication among devices happens through 
%an 
wireless medium on a shared asynchronous channel with packet loss due to noise and object's positioning. 
%All devices employ five types of messages 
%All devices exchange each other five types of messages. A broadcast control message to establish the cluster service. 
%There are five different types of messages exchanged by the objects. Control messages sent in \textit{broadcast} to form clusters. Messages from leaders to register themselves with the AP. \textit{Multicast} messages from AP to cluster leaders to disseminate a task. ACCEPT messages from cluster leaders to AP, informing of the acceptance of a task. Task dissemination from cluster leaders to their cluster members.
The system utilizes five different message types. The \textit{CapabilityDissemination} messages are sent in \textit{broadcast} to configure clusters. Leaders send \textit{LeaderRegister} messages to Access Point  (AP) to register themselves. The AP dispatches a task using \textit{TaskDispatch} messages and leaders accept them using \textit{TaskAccept}. Finally, leaders disseminate tasks to their cluster using \textit{LeaderToCluster} messages.
Further, all objects transmit in the same transmission channel in order to cluster. Moreover, the application layer can consume the sensing results using any application layer protocol, such as CoAP.

\subsubsection{\textbf{Task model}} Each task 
represents a sensing demand
that requires different sensing capabilities in order to be performed.  A task $T$ is a set $ \{ T_{id}, C, \tau, q \} $ on which $T_{id}$ is the unique id of the task; $C$ denotes the set of capabilities needed to complete the task, $\tau$ denotes the discrete time needed to complete it and $q$ the \textit{per-}cluster quorum to perform the task. Tasks are dispatched by the AP who also keeps track of their status which can be pending, dispatched or completed.
%Task T is dispatched through the Access Point (AP), which has a list of pending tasks.
%The Access Point (AP) has a list of pending tasks and is responsible for dispatching them.

%\section{CONTASKI Architecture}
\section{CONTASKI}
\label{sec:CONF}

%\noteblue{Discutir os nomes dos módulos em inglês}
%\al{Talvez repensarmos o formato da arquitetura}

\begin{figure}[ht]
	\centering
	\includegraphics[width=0.90\linewidth]{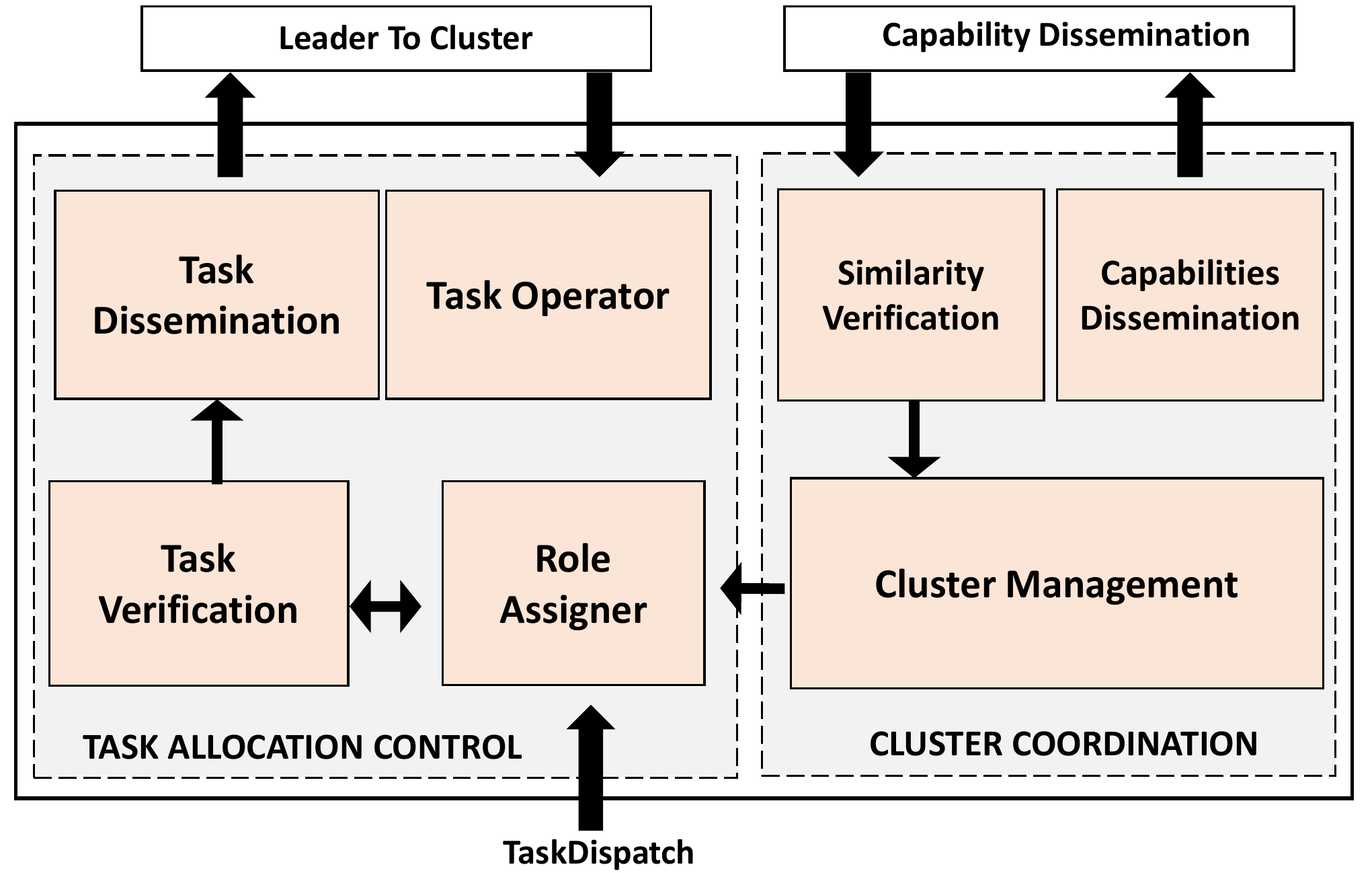}
	\caption{The CONTASKI Architecture}
	\label{fig:contaski-architecture}
\end{figure}

The CONTASKI architecture is comprised by two modules called: \textbf{Cluster Coordination} (\textbf{CCM}) and \textbf{Task Allocation Control} (\textbf{TACM}), as 
shown  
%seen 
in Fig.~\ref{fig:contaski-architecture}. 
%Both 
They act
%\as{work/play/act} jointly 
%in order 
to guarantee
%\as{both}
both task dissemination and allocation among %the
objects (nodes) of the network. The Cluster Coordination Module organizes the network in clusters and the Task Allocation Control manages the task dissemination among the network participants according to their capabilities.

The~(\textbf{CCM}) module 
%\textbf{Cluster Coordination Module} 
controls the creation and upkeep of the clusters. It evaluates the neighboring nodes using a similarity threshold of their capabilities in order to verify if they are apt to participate in the same cluster. Therefore, when it receives a \textit{CapabilityDissemination} message 
%\noteblue{tipo da mensagem},
it verifies the identification, capabilities and number of neighbors. CCM comprises three components, the component Capabilities Dissemination (CD), which is responsible for disseminating the~\textit{CapabilityDissemination} messages with the $Id$, capabilities and number of neighbors. The Similarity Verification (SV) component receives and interprets those messages exchanged among the nodes. Finally, Cluster Management (CM) component manages the cluster creation using the nodes' similarity. It is also responsible for leader selection.

The~(\textbf{TACM}) module
%~\textbf{Task Allocation Control Module} 
%coordinates
manages 
and dispatches the tasks to the nodes of the IIoT network in order to maximize and preserve nodes' resources. It comprises the components:  Task Verification (TV), Role Assigner (RA), Task Dissemination (TD) and Task Operator (TO). TV 
%component 
oversees which tasks are to be per\-for\-med and what capabilities are needed. RA monitors the tasks to be assigned to the nodes. It employs collaborative consensus to evaluate which tasks should be allocated according to each node's capability.
%Consensus involves the agreement and uniformity of opinions that the nodes establish through the exchange of information between them
DT dispatches the requested tasks honoring the capabilities of the nodes.
%in the network. 
TO is responsible for receiving and execution~of the tasks disseminated by the group leader. Thus, the task allocation becomes 
fare and balanced, and it does not overload the~nodes.

\subsection{Cluster configuration}
As the network size involves nodes with different 
capabilities, CCM arranges the network in clusters based on the leaders in order to create a  network infrastructure capable of allocating tasks.
%to task allocate.
%In this way, nodes initially  exchange \textit{CapabilityDissemination} messages to \as{notify} the $ID$, capabilities and its number of neighbors. 
%Initially, nodes begin cluster configuration exchanging \textit{CapabilityDissemination} messages that carries the $ID$, capabilities and number of neighbors of the sender.
%
Algorithm~\ref{alg:cluster-configuration}
describes %the operation of 
the cluster coordination module. At first, each node sends a \textit{CapabilityDissemination} message in order to announce its $ID$, capabilities ($Capabilities$), and neighborhood size. There is a random send interval to prevent simultaneous transmissions.
When receiving a \textit{CapabilityDissemination} message, receiver updates its neighbors structure ($NeighList$), alongside with their capabilities ($NeighCapabilities$). The similarity verification takes into account those structures, being calculated using cosine similarity. The neighbor can join the cluster when its similarity is within the threshold. This update procedure happens dynamically in every node, ensuring each node maintain its neighbor and cluster structure updated.
\begin{algorithm}[ht]
	\relsize{-1.9}
	\textbf{procedure} \textsc{SendCapabilityMessage}{}\\
	\sep $Broadcast(Id, MyCapabilities,N_{neigh})$ \\
	\sep $WaitInterval()$ \\
	\textbf{end procedure}\\
	\BlankLine
	\textbf{procedure} \textsc{ReceiveCapMessage}{($Id, Capabilities, N_{neigh}$)} \\
	\sep $NeighList \leftarrow NeighList \cup Id$ \\
	\sep $NeighCapabilities[Id] \leftarrow Capabilities$ \\
	\sep $NeighSize[Id] \leftarrow N_{neigh}$ \\
	\sep $SimilarityCalculation(NeighCapabilities, Id)$ \\
	\textbf{end procedure}
	
	\BlankLine
	
	\textbf{procedure} \textsc{SimilarityCalculation}{$(NeighCapabilities, Id)$} \\
	\sep $NCapabilities \leftarrow NeighCapabilities[Id]$ \\
	\sep $sim = \frac{|MyCapabilities \cap NCapabilities|}{\sqrt{|MyCapabilities|*|NCapabilities|}}$ \\
	\sep \If{$sim \geq Treshold$}{\sep $cluster \leftarrow cluster \cup Id$}
	\sep $SelectLeader(NeighCapabilities)$ \\
	\textbf{end procedure}
	
	\BlankLine
	
	\textbf{procedure} \textsc{SelectLeader}{$(NeighCapabilities)$} \\
	\sep \For{$neigh \in NeighList$}{
		\sep \If{NeighSize[neigh] is the greatest one}{
			\sep $ClusterLeader \leftarrow neigh$ \\
			\sep \If{$ClusterLeader = SId$}{
				\sep $SendMessageToAP(SId)$
			}
		}
	}
	\textbf{end procedure}
	
	\caption{Cluster configuration}
	\label{alg:cluster-configuration}
\end{algorithm}
%\vspace{-0.07cm}

\begin{comment}
\begin{figure}[h]
	\centering
	\includegraphics[width=0.8\linewidth]{images/similaridade}
	\caption{Similarity scale}
	\label{fig:similarity-rule}
\end{figure}

The similarity value
%\as{value/grade/index} 
is computed based on the capabilities between two nodes and varies from 0.65 to 1. Closer to 1, more similar the two nodes are. Fig.~\ref{fig:similarity-rule} shows the similarity in relational level in $S_1$ = Weak, $S_2$ = Medium and $S_3$ = Strong. This scale varies according to the pre-established capabilities before the network is deployed.
\end{comment}

The leader selection process
%\as{process/stages/step} 
takes into account~both the~number of neighbors and individual capabilities to choose the cluster leader. After that, the selected leader registers itself with the AP %.~\cm{This process}
%\as{Leader selection/This process} 
%in order 
to guarantee the communication between nodes and the AP and a better hierarchical network organization.% in the network.

%Similarity value is computed as follows $sim(ob_1,ob_2) = \frac{|C_{ob1} \cap C_{ob2}|}{ \sqrt{|C_{ob1}|*|C_{ob2}|} }$, being based on~\cite{Chen2016}.

%\begin{comment}
Equation~\ref{eq:similarity}
%\as{verifies/
computes the similarity value
%\as{value/index/grade of} 
%similarity 
between two nodes capabilities, being based on~\cite{Chen2016}.
%\end{comment}
The similarity takes into account the node's own capabilities ($C_{ob1}$) and the neighbor's capabilities ($C_{ob2}$). The upper part calculates the norm of the vector that represents the intersection between the capabilities. The bottom part takes the square root of the multiplication of the norm of each capability vector.

\begin{equation}
\label{eq:similarity}
sim(ob_1,ob_2) = \frac{|C_{ob1} \cap C_{ob2}|}{ \sqrt{|C_{ob1}|*|C_{ob2}|} }
\end{equation}

\subsection{Task Allocation}
The tasks are made available through the AP, where a list is kept with a set of tasks to be performed. The tasks are sent by messages directed to the clusters leaders. In order to identify the leaders, the AP monitors the \textit{LeaderRegister} messages and keeps a leader list updated. \textbf{TACM} relies on a network infrastructure established by the cluster configuration, so that it runs guaranteeing resource maximization, i.e., allowing the task dissemination according to 
the capabilities of each cluster.

\begin{algorithm}[ht]
\relsize{-1.9}

\textbf{procedure} \textsc{SendTask}{} \\
\sep \For{$task \in TaskList$}{
	\sep $Multicast(ClusterLeaders, T_{id},C,\tau, q)$ \\
	\sep \If{$WaitConfirmation()$}{
		\sep $TaskList \leftarrow TaskList - {task}$
	}
}
\textbf{end procedure}\\

\BlankLine

\textbf{procedure} \textsc{ReceiveTask}{$(T_{id},C,\tau, q)$} \\
\sep \If{
	%not busy \textbf{and}
	$C \subseteq MyCapabilities$
	\textbf{and}  $|cluster|$ $\geq q$}{
	\sep $ReplyToAP(SId, T_{id}, ``TaskAccept")$ \\
	\sep $BroadcastToCluster(T_{id}, \tau)$ \\
}
\textbf{end procedure}\\

\caption{Task Allocation}
\label{alg:task-allocation}
\end{algorithm}

Algorithm~\ref{alg:task-allocation} describes the task allocation control between nodes, how the entities relate to themselves in performing tasks dispatched by the AP.
Initially, the AP holds a list of pending tasks ($TaskList$).
%In order to dispatch a task,
When dispatching a task, the AP selects a pending task from the list and sends a \textit{TaskDispatch} message to the cluster leaders announcing the task $T$ to be executed.
After the dispatch, the AP waits $WaitConfirmation()$ until it receives the confirmation (\textit{TaskAccept} messages) of which cluster leaders can perform that task.
When at least one confirmation is received, the AP moves the task out of the pending task list. When the leaders receive the task $T$, they verify if their own capabilities $MyCapabilities$ are compatible with the capabilities $C$ needed to perform the task, and if the number of nodes in the cluster is greater or equal the quorum $q$ needed.
Case they meet the criteria, the cluster leader confirm to the AP with a \textit{TaskAccept} message that it will perform the task and disseminates the task to the cluster.
%Leaders that do not confirm, cannot perform that task.
Case the cluster members cannot realize %perform 
the task, the leader doesn't confirm this task with the AP.
As the allocation management acts in a dynamic and collaborative way, together with the cluster configuration, participants reach a better task distribution among them. Nonetheless, it is emphasized that the distributed operation within the network depends on consensual collaboration between all participants.

\subsection{Operation}
\begin{comment}

%All nodes participate in the cluster configuration
\al{Este parágrafo abaixo parece redundante e precisamos  verificar se ele fica} 

\as{
Cluster configuration 
happens %takes place 
dynamically in each node. Interactions between the nodes takes place in space and time, 
%and nodes within the transmitter range exchange messages. 
therefore messages are sent and received by node withing the transmitter range. 
Each node announces (broadcasts) its identifier, capabilities and number of neighbors by using a \textit{CapabilityDissemination} message. When receiving this message, the receiver node verifies all message fields and computes the similarity value. Case the value is within the threshold, %the other node 
the sender node joins the cluster.}
\end{comment}

\begin{figure}[ht!]
	\centering
	\includegraphics[width=0.95\linewidth]{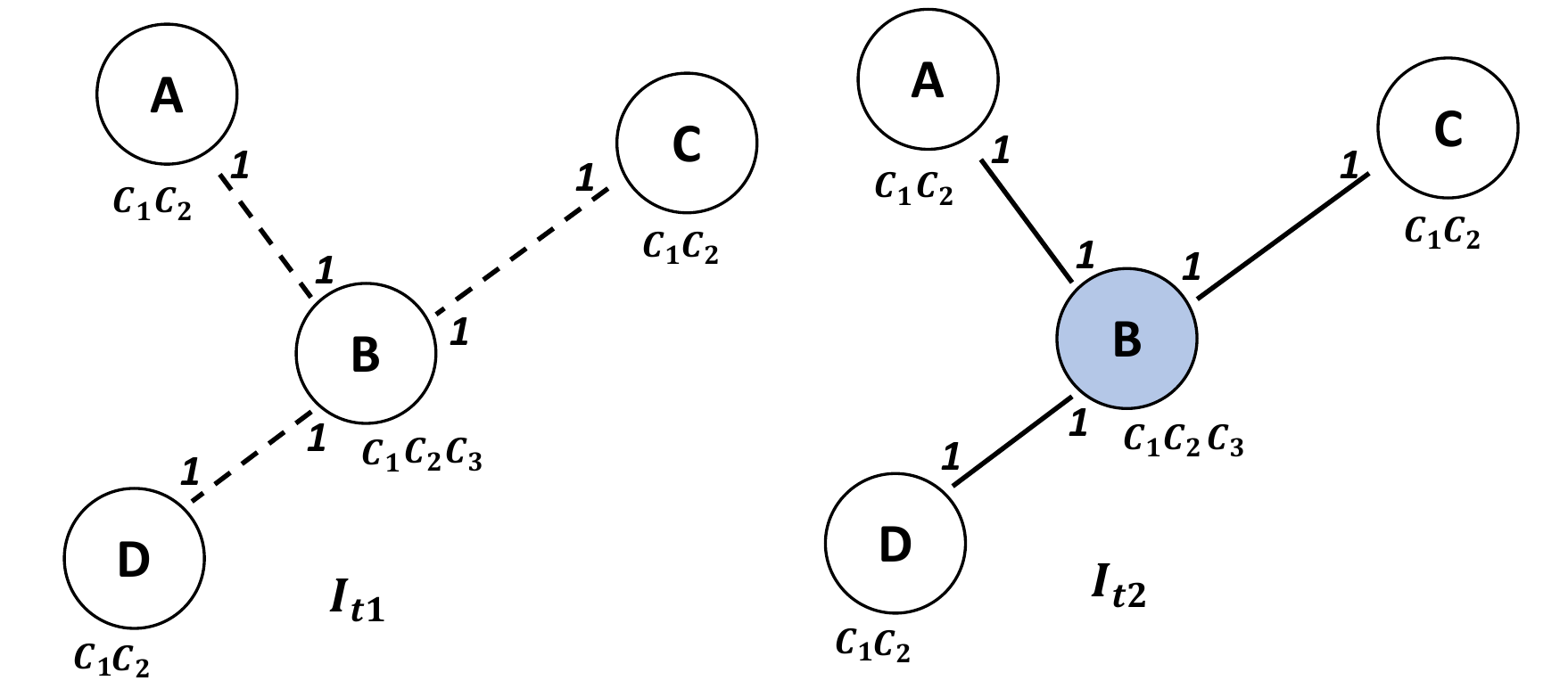}
	\caption{Clustering procedure}
	\label{fig:cluster}
\end{figure}

Fig.~\ref{fig:cluster} exemplifies the cluster configuration and leader selection phase. Dashed edges indicate nodes within each other's transmission range and thus apt to exchange control messages. Each node holds its identifier, set of capabilities and number of neighbors. Each time instant $I_{t}$ corresponds to message exchange for cluster configuration and leader selection. On instant $I_{t1}$ the 
group  of nodes (\textbf{A}, \textbf{B}, \textbf{C}, and \textbf{D})
share capability information and, then, compute the similarity according to Equation~\ref{eq:similarity}. Nodes compute the following similarities: $sim_{A, B}=\frac{2}{\sqrt{2*3}}=0,81$, $sim_{A,C}=sim_{A,D}=\frac{2}{ \sqrt{2*2}}=1$, $sim_{B, C}=sim_{B, D}=\frac{2}{ \sqrt{2*3}}=0,81$, $sim_{C, D}=\frac{2}{\sqrt{2*2}}=1$. The minimum similarity is $0.81$ and maximum $1$. 
%According to the similarity scale in Figure~\ref{fig:similarity-rule}, the minimum similarity corresponds to $S_{2}$ = Medium. 
Therefore, all nodes within that interval are clustered and that cluster represents nodes with capabilities $C_{1} C_{2}$.

\vspace{-0.3cm}

On instant $I_{t2}$, the leader selection happens according to the procedure $SeectLeader$ in Algorithm~\ref{alg:cluster-configuration}. 
Since node~\textbf{B}
scores the highest number of 
neighbors, it is selected as leader. With the Cluster Coordinator operating in this manner, each node maintains its neighbors' information updated through message exchange. The cluster structure 
%That structure \noteblue{Que estrutura?} 
determines which nodes in the spatial neighborhood are seen as members of the same cluster.
%, in addition to ensuring a better scalability to the network, since a hierarchy based on leaders helps in the quality of the information transferred.
It also eases the dissemination of tasks among nodes, because every leader represents the cluster itself.

The \textbf{TACM} acts considering the clusters are already configured. Once the leader is chosen, it sends a \textit{LeaderRegister} message to the AP informing that it represents that region of the network. The AP has the list of tasks to be performed and that are allocated according to the application's priority. Each task requires a minimal capability set to be performed and can run for an arbitrary time. Thus, when receiving a task, the leader verifies the cluster capabilities to make 
%are compatible with ones required for 
the task. In case positive, the leader sends a \textit{TaskAccept} message to the AP informing that the cluster will perform the task.

\begin{figure}[ht]
	\centering
	\includegraphics[width=1\linewidth]{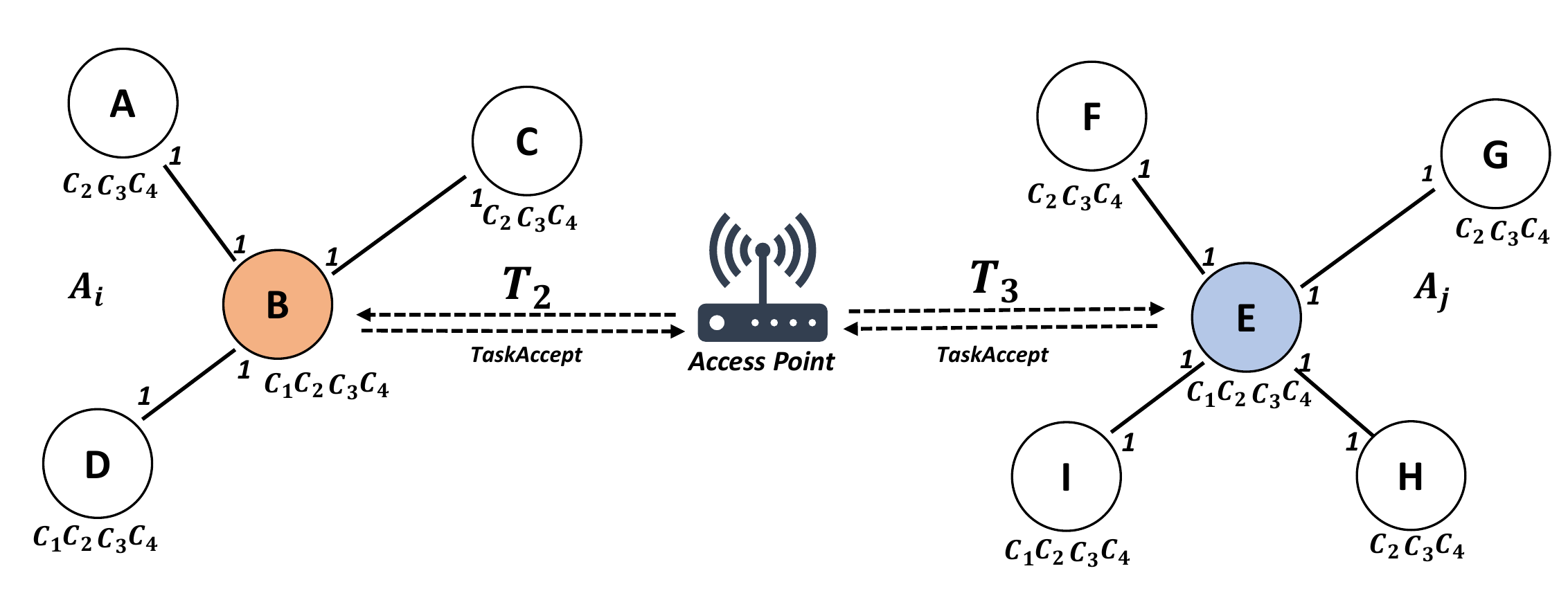}
	\caption{Task allocation procedure}
	\label{fig:task-allocation}
\end{figure}

Fig.~\ref{fig:task-allocation} depicts the CONSTASKI operation in the task allocation process between AP and nodes. In the example~there are two clusters $A_i$ and $A_j$, comprised of nodes (\textbf{A}, \textbf{B}, \textbf{C} and \textbf{D}) and (\textbf{E}, \textbf{F}, \textbf{G}, \textbf{H} and \textbf{I}), respectively. Each cluster keeps a leader, $A_i = \textbf{B}$ and $A_j = \textbf{E}$, 
responsible for directing communication with the AP and 
disseminating tasks among cluster members. 
In this way, the AP verifies its set of tasks and sends the  \textit{TaskDispatch} message with the task $T$ to cluster leaders.  Thus, Leaders \textbf{B} and \textbf{E} 
will answer with the \textit{TaskAccept} message, confirming that their cluster can carry out that task. But this is a partial view of the network, in the occurrence of a cluster that cannot perform that task, it is sufficient to not send the~\textit{TaskAccept} message. In this way, all tasks are allocated according to the node capabilities, and nodes that cannot perform that task wait for a new task round.

\vspace{-0.4cm}

\section{Analysis}
\label{sec:ana}

This section shows a performance evaluation of the CONTASKI mechanism in order to assess the gains to the management of all task. We implemented CONTASKI in NS3-simulator, version 3.29, and make all simulations taking into account an IIoT scenario similar to a manufacture industry environment. These  objects  may  vary  according  to  the type of  industry  evaluated  and  their  type  of  function  within  the environment. The capabilities found at each object follow the sensing functions %described 
in~\cite{xu2019big}, 
%The roles/capabilities found at/realized by each object follow the sensing functions described/found  in~\cite{haverkort2017smart}.
which consist of %the following sensing: 
{\it temperature}, {\it humidity}, {\it presence}, {\it light}, {\it machine status},
%{\it machinery synchronization},
{\it pressure} and {\it reservoir level}.
%The choice of node capabilities is based on the sensing functions that an IoT object can perform and follows the functions %described 
%in~\cite{haverkort2017smart}.
The industrial scenario set up
%The evaluation industrial scenario 
%consisted of 
involved 50, 75 and 100 nodes evenly %randomly
distributed in a rectangular area of~\textit{200 x 200} meters %in order 
to create clusters that fit the task distribution, considering nodes with different sensing capabilities.
%~\cm{and this allows building clusters that fit the task distribution, considering nodes with different sensing capabilities.}
%\as{e isso permite construir agrupamentos que atendam a distribuição da tarefa, considerando nós com diversas capacidades de sensoriamento}. \as{O tempo de operação foi...}
%and 

%running over \textit{900s}
%The time operating for~\textit{900s},~\cm{this time is related to sensing tasks and warm-up, each task carries out for~\textit{60} seconds}. With a transmission radius of \textit{100m}. 
%\al{The total system operation time is~\textit{900s}. During the initial~\textit{150s}
%\al{The total system operation time spends \textit{900s}

\begin{comment}
\begin{figure*}[ht!]
    %\lefting
    \centering   
    %\hspace{-1cm}
  
    \includegraphics[width=0.32\linewidth]{images/funcionamento_50.pdf}
    \includegraphics[width=0.32\linewidth]{images/funcionamento_75.pdf}
    \includegraphics[width=0.32\linewidth]{images/funcionamento_100.pdf}
    \caption{Number of clustering %establishment 
    over time}
      \vspace{-0.07cm}
      \label{Fig:fun}
\end{figure*}
\end{comment}
The system operates for~\textit{800s} and over the initial~\textit{150s} all nodes warm-up exchanging their capabilities through messages, followed by the similarity calculation and leader register. The AP dispatches 10 tasks from~\textit{150s} to up~\textit{740s}.~Clusters perform each task for~\textit{60s}.  In addition, each task has~a~random capabilities set, and all tasks require 
%have 
at least the temperature, humidity and presence capability set. The final capability set has up to 4 other capabilities between light, synchronization, pressure and reservoir level capabilities.
Nodes communicate through the IPv6 protocol over an~\textit{ad-hoc} IEEE 802.15.4 network.
%Further, all packets can be lost in the transmission and we use up to five retransmissions in case of errors or lost packets.
We also consider that nodes do not present communication failures and, in order to account for packet loss due to interference and bottleneck, we added a delay of 2ms for messages exchanged among nodes.
%We added a time of~\textit{2} milliseconds for messages exchanged among nodes to avoid packet loss and~bottleneck.

The tasks  to be performed are directed to leaders through the AP, always available, located in the center of the network and equipped with a strong internet signal to reach all nodes. The similarity parameter varies from 0.65 to 1, closer to 1, higher is the similarity between two nodes. The 
%network %acting 
nodes are set up with 
%are configured with different 
varied capabilities. We assess 
%the performance and efficiency  of 
CONSTASKI using metrics based on~\cite{khalil2019new}:~\textbf{number of clusters (NC)},~\textbf{number of allocated tasks (NAT)},
%~\textbf{number of unallocated tasks},
~\textbf{clusters apt to perform tasks (CPT)}, ~\textbf{clusters inapt to perform tasks (CIT)} and,~\textbf{latency of accept time (LAT)}.
%and~\textbf {number of rounds (NR)}. 
The results obtained in all simulations correspond to the average of 35 simulations with 95\% of confidence interval.
A comparative analysis with the work of~\cite{khalil2019new} was not carried out due to scenario incompatibility since they divide the network in two clusters and 
perform only one type~of~task.

\subsection{Results} % Evaluation}

\begin{figure}[h]
    %\lefting
    \centering   
    %\hspace{-1cm}
    \includegraphics[width=0.5\linewidth]{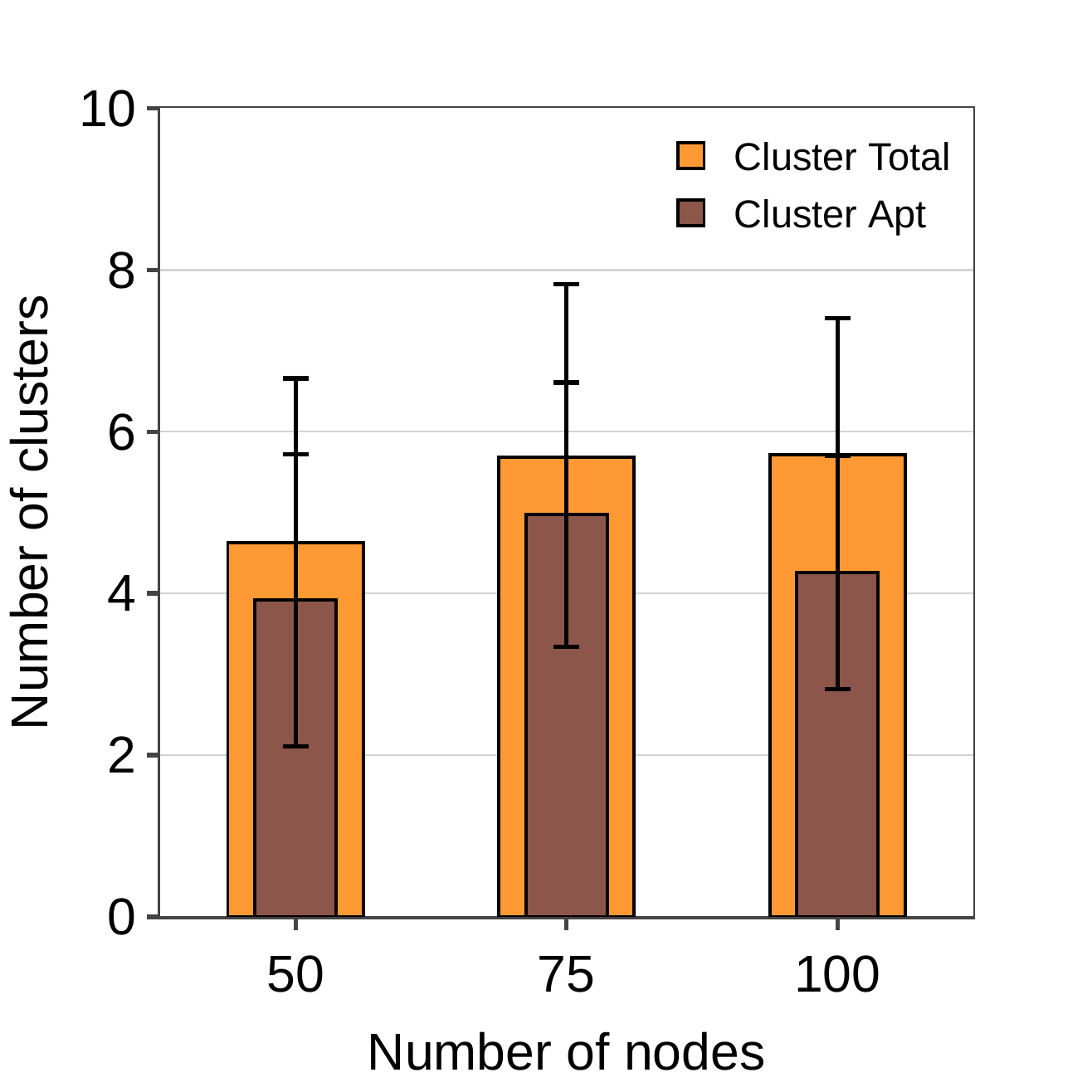}
    \vspace{-0.08cm}
    \caption{%Established/
    Total clusters %X
      versus clusters apt to perform tasks}%establishment }
      % Cluster total/Full
      % cluster apt 
      \label{fig:apt-inapt}
\end{figure}

\begin{figure}[h]
    %\lefting
    \centering   
    %\hspace{-1cm}
    \includegraphics[width=0.8\linewidth]{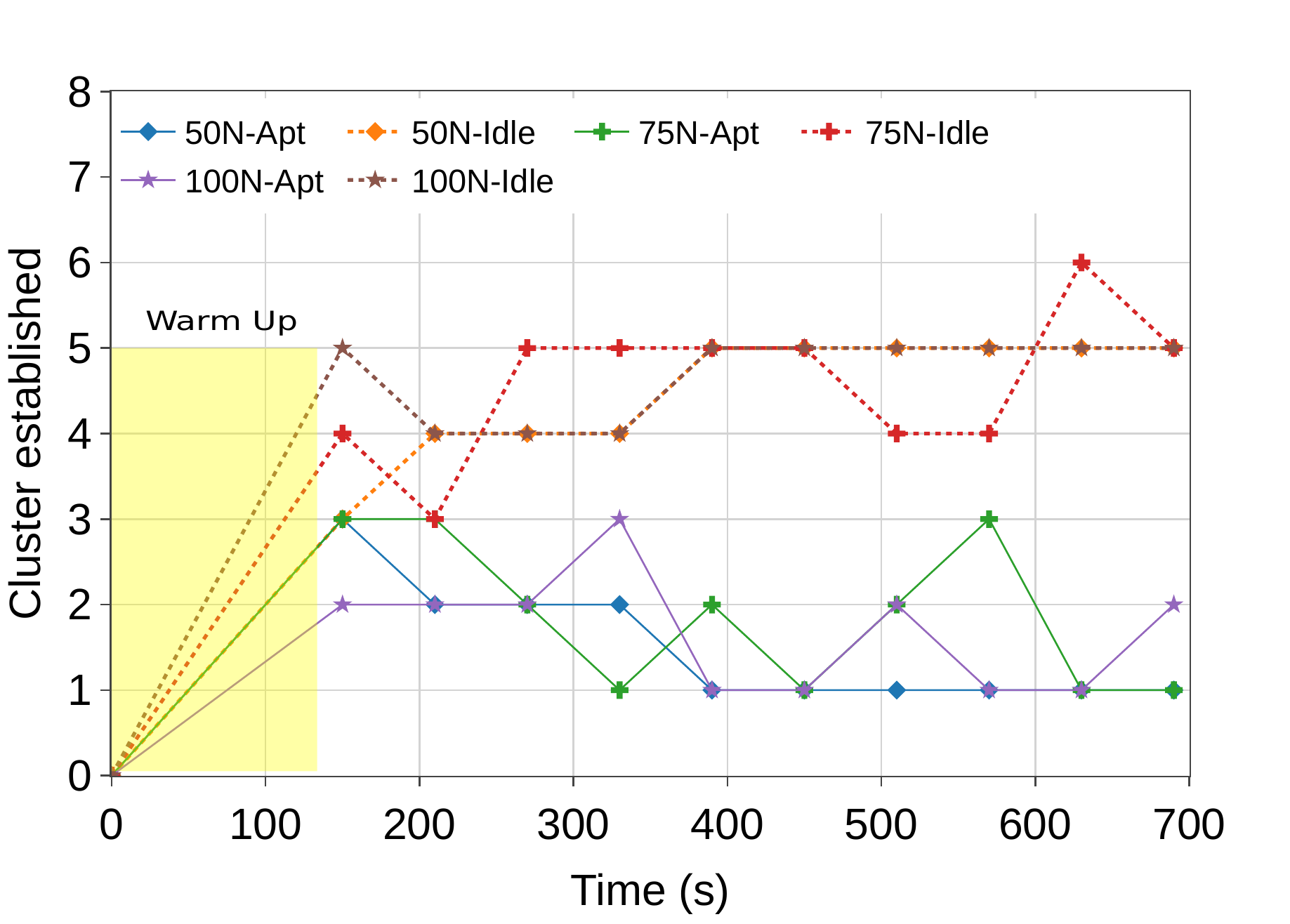}
     \vspace{-0.08cm}
    \caption{Number of clusters over time}
      \label{fig:funcionamento}
\end{figure}

Fig.~\ref{fig:apt-inapt} shows the CONTASKI performance for supporting the cluster formation.  
%The CONTASKI 
%performance for cluster formation is~shown~in Fig.~\ref{fig:apt-inapt}. 
\textbf{NC} has a direct relation to the similarity among nodes that takes into account each capability set and the capabilities set of their neighbors. Clusters are labelled as apt and inapt and the differentiation happens on each task dispatch. Apt clusters can perform the dispatched task, otherwise, the cluster is considered ``inapt''. Also, considering the static scenario and transmission range of the nodes, they form clusters with a non-deterministic number of participants. CONTASKI was able to have an average of CPT with one or at most two points of difference to NC.
Fig.~\ref{fig:funcionamento} shows the CPT and CIT
%number of apt and inapt cluster 
in each task dispatch. Each point 
means %represents
a task being dispatched, dotted lines mean
%represent 
inapt clusters and solid lines, apt ones. All 10 tasks were dispatched and performed up to \textit{700s}.
It is seen that %the number of 
inapt clusters is significantly higher than the apt ones.
Despite that, all tasks have at least one apt cluster performing them.
%This situation occurs given that each cluster can be counted twice since the differentiation between apt and inapt is done in each task dispatch.
%Furthermore, an inapt cluster saves its energy resources for the next task round.
%Given that the cluster leader compares its capabilities to those required by the task, an idle cluster could not perform the task.

\vspace{-0.4cm}
%\as{A Fig. 4 com um valor máximo no eixo Y de 12 ficará num formato mais proporcional, isto é formato de paisagem (horizontal) no lugar de vertical (retrato) }

\vspace{-0.8cm}

\begin{figure}[ht!]
    %\lefting
    \centering   
    %\hspace{-1cm}
    \includegraphics[width=1\linewidth]{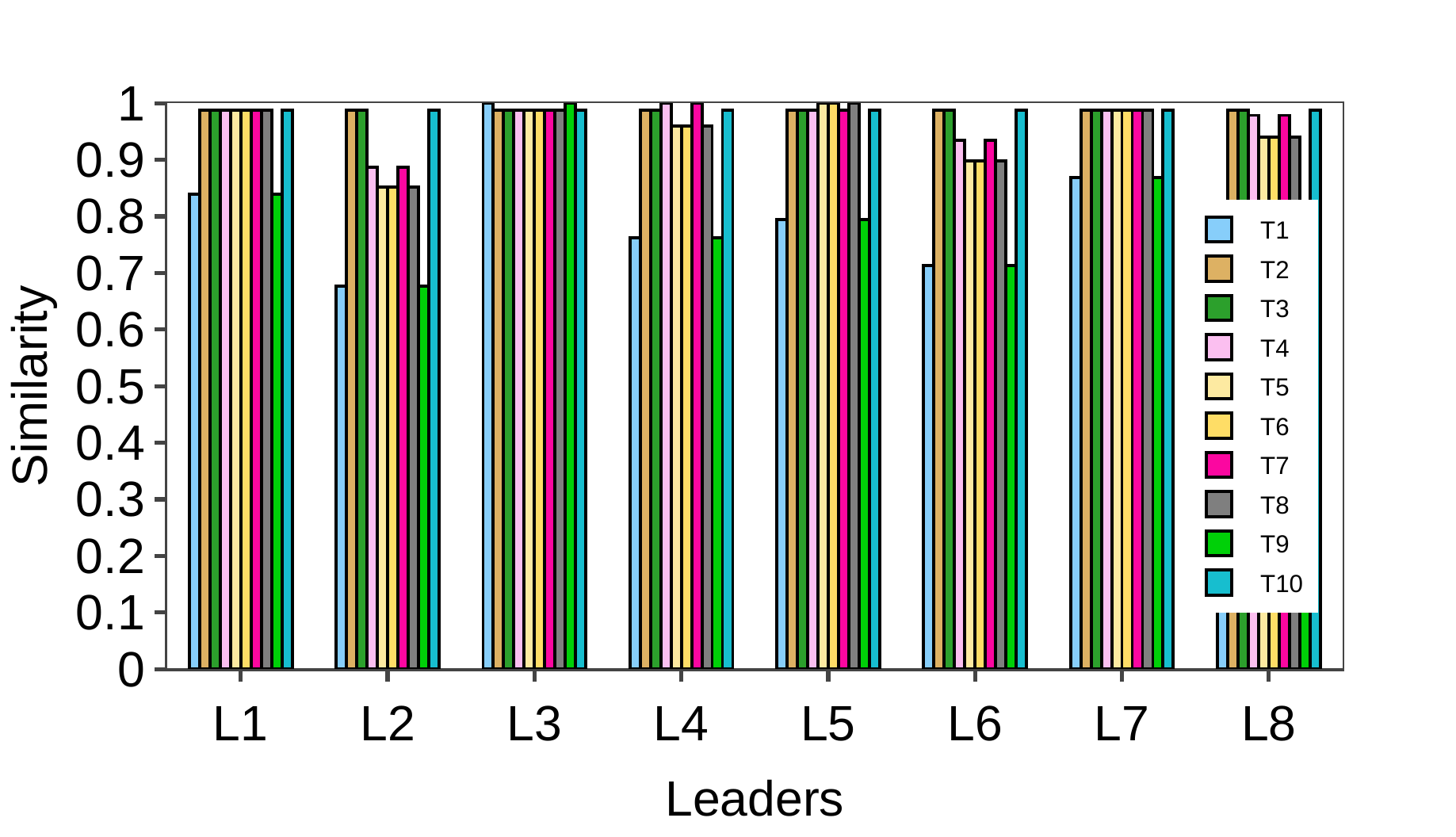}
     \vspace{-0.08cm}
    \caption{Similarity between task and leaders}
      \label{fig:task-leader-similarity}
\end{figure}

\begin{figure}[ht!]
    %\lefting
    \centering   
    %\hspace{-1cm}
     \includegraphics[width=0.45\linewidth]{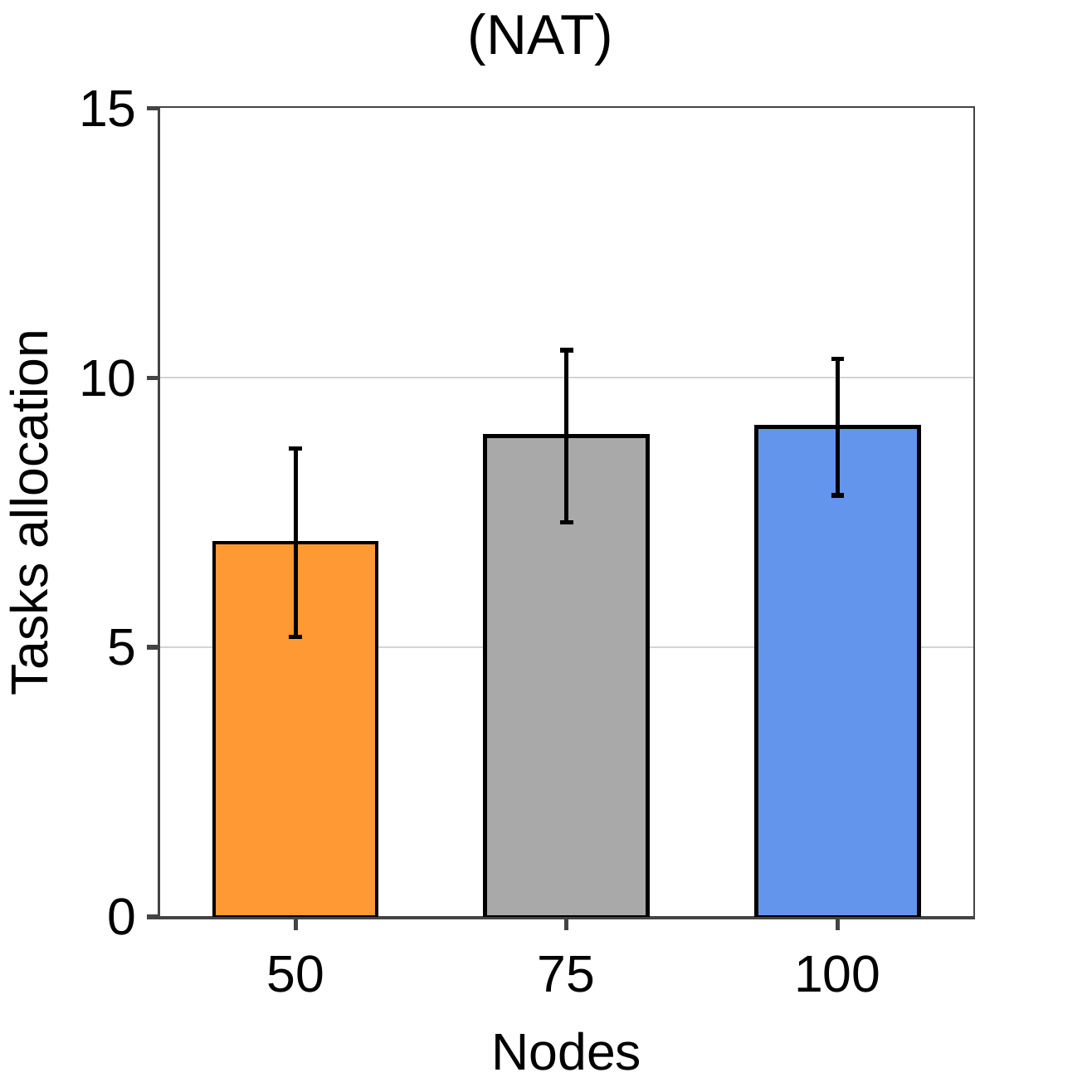}
    \includegraphics[width=0.45\linewidth]{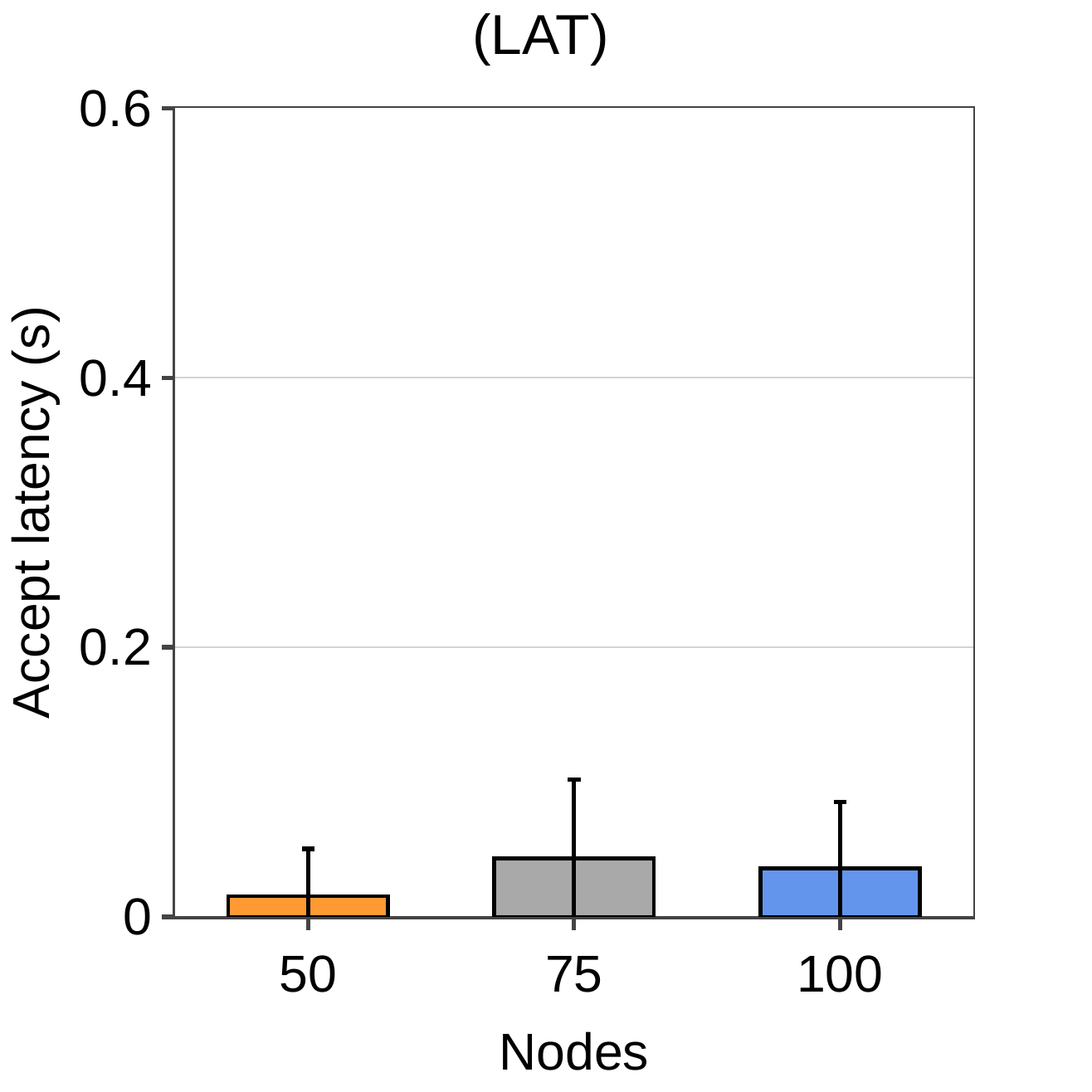}
    \caption{Task allocations and latency in task acceptance}
   \vspace{-0.07cm}
      \label{fig:tasks-latency}
\end{figure}

Fig.~\ref{fig:task-leader-similarity} shows
%\as{ratio of ?}
the similarity correlation between tasks and cluster leaders in the seventh round with 50 nodes. This round exhibited nodes' capabilities similar to task's capabilities and higher amount of leaders. Colored bars means each dispatched task. Leaders computes similarity between its capabilities and tasks' demanded capabilities, and accept the task whose similarity equals to one. Fig.~\ref{fig:tasks-latency} \textbf{(NAT)} shows the amount of tasks dispatched by the AP, which is at most 10. Each task takes~\textit{60s} and can be realized by more than one cluster. Clusters in a scenario with 75 and 100 nodes were able to perform over 90\% of dispatched tasks.
The distortion with 50 nodes is due to the small number of nodes and the random capabilities assigned to them. The randomization make it unable to guarantee that nodes have compatible capabilities in relation to the capabilities needed by the tasks. CONTASKI was able to reach a low LAT, that means the difference between the task dispatch time and the last accept time as seen by the AP. This is related to the hierarchical infrastructure based on cluster leaders. Since the capabilities of leaders  are compatible with the cluster participants, the leader is responsible for evaluating if the cluster can perform the task and inform the AP about the acceptance. LAT varies according to the number of existing clusters and apt clusters for a given task e.g. with 100 nodes the achieved latency was \textit{35ms}, being smaller than one with 75 nodes, that achieved \textit{43ms}. The smallest latency was \textit{15ms}~with~50~nodes.

\vspace{-0.5cm}

\vspace{-0.4cm}

\vspace{-0.3cm}
\section{Conclusion}
\label{sec:con}
This work presented CONTASKI for allocating tasks on devices in an IIoT network. It organizes the network into clusters based on the similarity of the capabilities of the devices and the capabilities of the neighboring devices. The mechanism applies the collaborative consensus to manage and distribute tasks between the clusters, considering the capacities they inform. These strategies manage to keep the network organized hierarchically allowing the participants to act according to their capabilities. Results show the effectiveness of CONTASKI 
%in the allocation of tasks 
by considering the relationship between capabilities and tasks to be performed by the  devices. Also, the high number of suitable clusters ensures better use of resources, increasing the quality of the information made available by the devices.
%nodes. 
As future work, we intend to evaluate the allocation of multiple simultaneous tasks between nodes, different contexts of IoT networks with different types of mobility. %Besides, an assessment of energy consumption and comparison with other task allocation mechanisms.

\vspace{0.2cm}
\textbf{Acknowledgments} - The authors would like to thank  RNP HealthSense Project (Grant no. 99/2017 in Brazil).

\bibliographystyle{IEEEtran}
\bibliography{sbc-template.bib}

\end{document}